\documentclass[12pt]{article}
\usepackage{a4,latexsym,graphicx,times,color,psfrag}

\usepackage{natbib}

\title{Are large two-dimensional clusters of perimeter-minimizing bubbles of equal-area hexagonal or circular?}
\author{{\bf S.J. Cox} \\ 
Institute of Mathematics and Physics, Aberystwyth University, SY23 3BZ, UK\\
{\bf F. Morgan}\\
Department of Mathematics and Statistics, Williams College, \\
Williamstown, MA 01267, USA\\
{\bf F. Graner}\\
Institut Curie, 26 rue d'Ulm, 75248 Paris Cedex 05, France,\\
and\\ 
Mati\`ere et Syst\`emes Complexes, Universit\'e Paris Diderot, CNRS UMR 7057,\\
  10 rue Alice Domon et L\'eonie Duquet, 75205 Paris Cedex 13, France
}

\date{June 2012}

\begin{document}

\maketitle

\begin{abstract}
A computer study of clusters of up to 200,000 
equal-area bubbles shows for the first time that 
rounding conjectured optimal hexagonal planar 
soap bubble clusters reduces perimeter.
\end{abstract}

\section{Introduction}

A single two-dimensional soap bubble minimizes its perimeter at fixed area \cite[]{WeaireH99}. When two bubbles meet, they can reduce the total (internal + external) perimeter of this nascent cluster by sharing an edge. The least perimeter way to fill the plane with bubbles of equal-area is to tile it with regular hexagons \cite[]{hales01}. Thus we expect that the least perimeter arrangement of a finite cluster of $N$ bubbles will consist of hexagons close to the centre, with any non-hexagonal bubbles (defects) close to the periphery. \citet{Coxg03} conjectured, on the basis of computer experiments on ``perfect'' clusters with $N$ a hexagonal number (of the form $3i^2 + 3i+1$) and a few other cases, that the shape of the periphery itself should also be hexagonal. \citet[][Figure 13.1.4]{morgan4th}, on the other hand, recognised that rounding the periphery of the cluster should reduce the total perimeter at large $N$. Here, we explore the competition between keeping the hexagonal shape of as many bubbles as possible, and reducing the total periphery of the cluster by making it circular.

\section{Methods}
\label{sec:methods}

\begin{figure}
\centerline{
(a)
\includegraphics[width=0.22\textwidth]{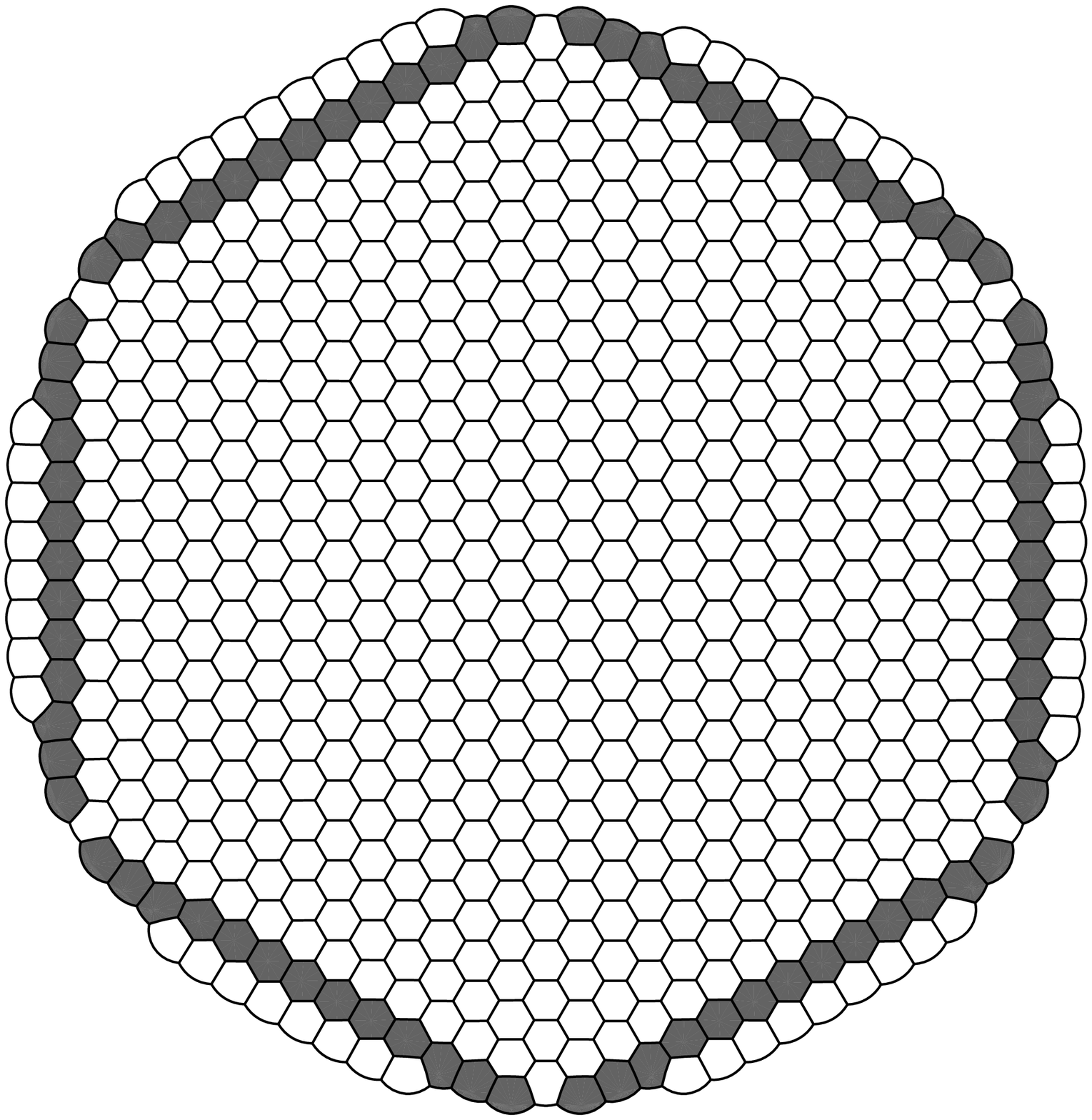}
(b)
\includegraphics[width=0.22\textwidth]{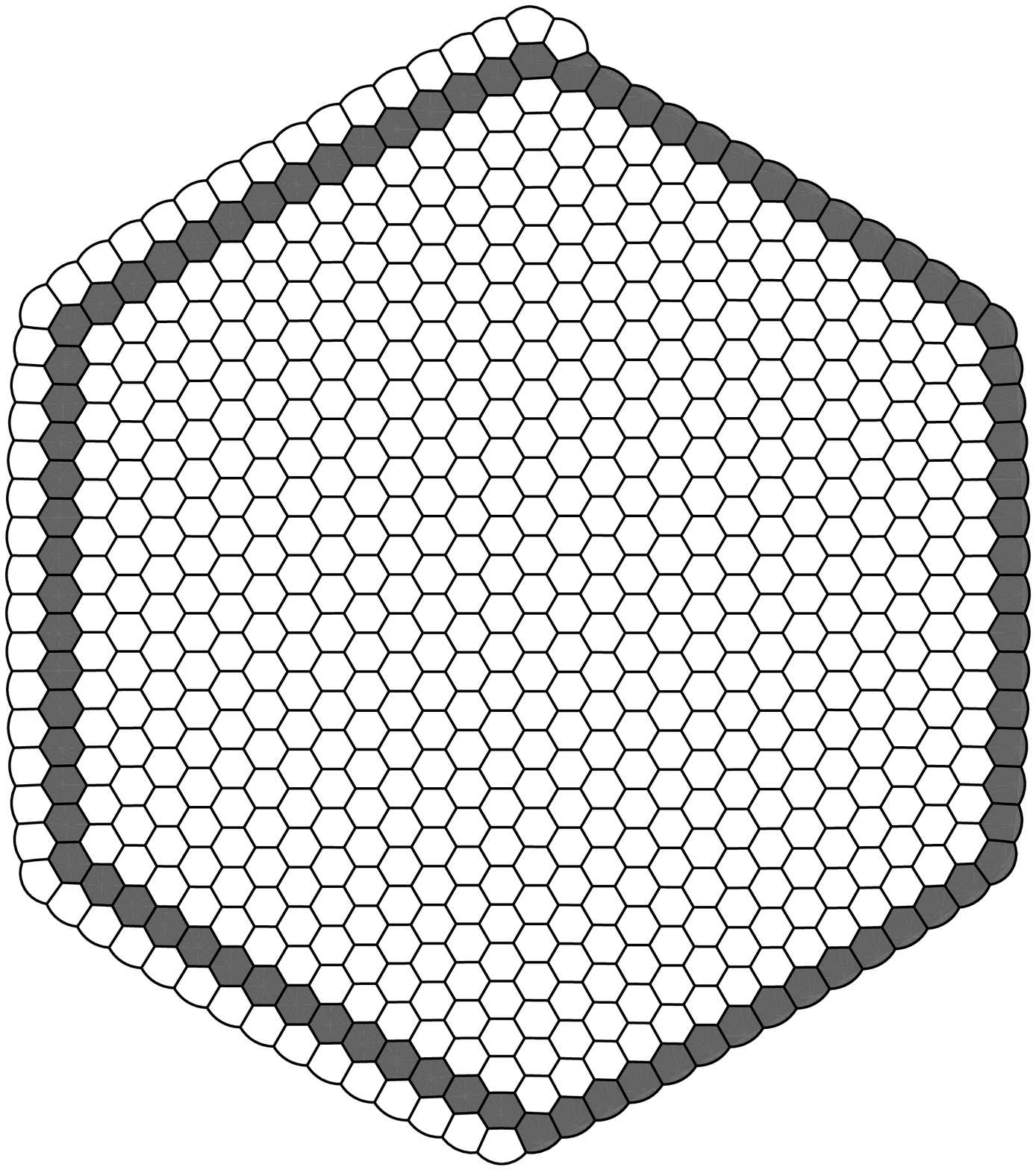}
(c)
\includegraphics[width=0.22\textwidth]{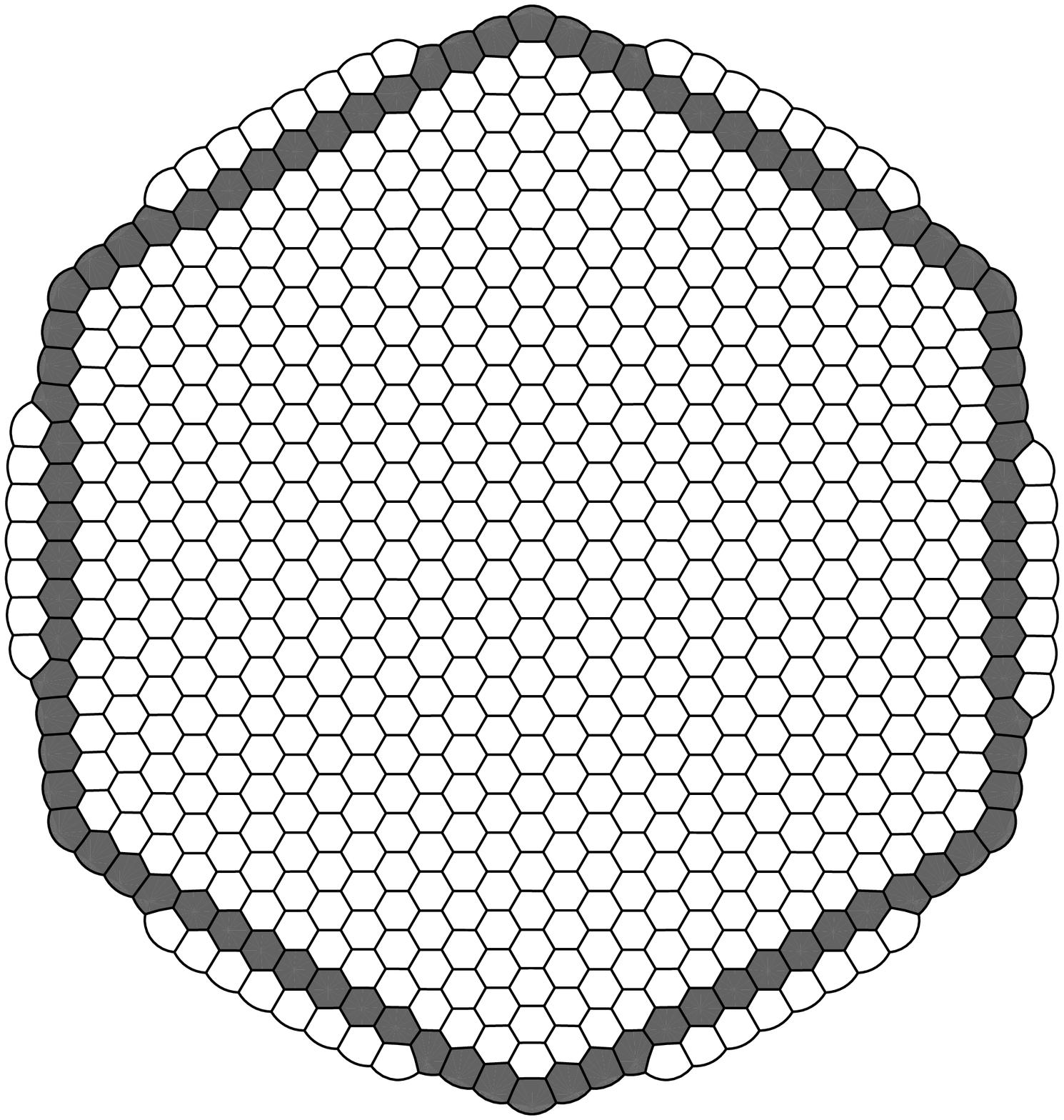}
(d)
\includegraphics[width=0.22\textwidth]{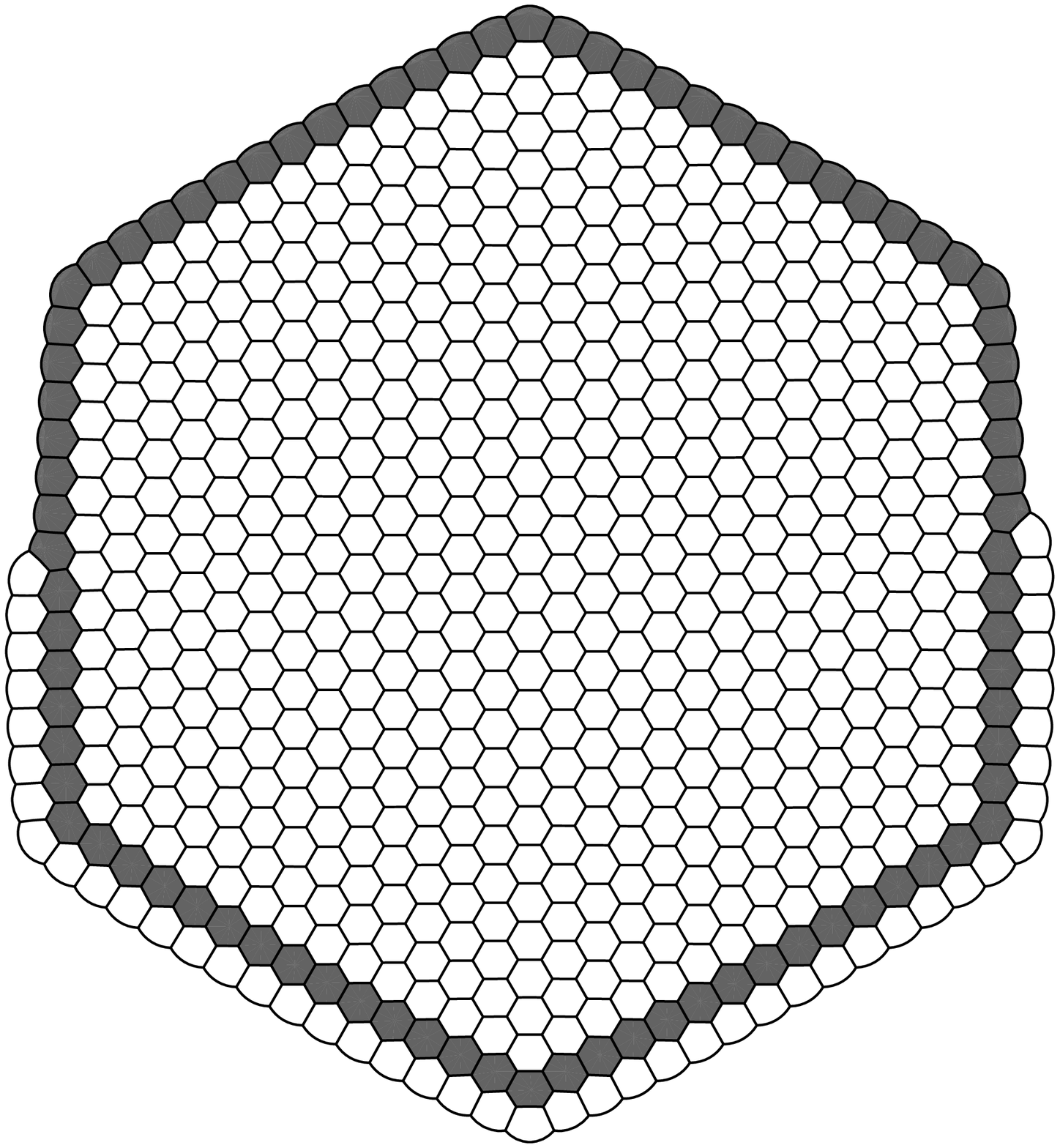}
}
\caption{Different equal-area clusters of $N=677$ bubbles, with the penultimate shell of bubbles shaded. (a) Circular $P_{circ}=2112.097$. The hybrid cluster is the same for this $N$. (b) Spiral hexagonal $P_{hexs}=2112.168$. (c) Corner hexagonal $P_{hexc}=2112.455$. (d) Topdown hexagonal $P_{hext} =2112.049$. For this $N$, the topdown hexagonal cluster is best.}
\label{fig:hexpics}
\end{figure}

We consider circular cluster, hexagonal clusters and hybrid clusters (defined below) of $N$ bubbles. Here we investigate $N$ up to $1000$ and $N$ a hexagonal number less than $11,000$. That is, we construct a cluster with hexagons in the bulk and the periphery of the required shape in Surface Evolver \cite[]{brakke92}, set all bubble areas to be equal (to $A_0 = 3\sqrt{3}/2$, so that edge lengths are close to unity) and seek a local minimum of the perimeter $E$, as described by \cite{Coxg03}. In practice, we start from a hexagonal cluster (e.g. with $N=1027$) and eliminate one bubble at a time, using one of the protocols described below and illustrated in figure \ref{fig:hexpics}:

\noindent{\bf Circular cluster}: The bubble whose centre (defined as the average of the positions of its vertices) is farthest from the centre of the cluster (defined in the same way) is eliminated.

\noindent{\bf Hexagonal cluster}: We take {\em hexagonal} to mean that all shells of hexagons except the outer one must be complete. We consider three processes of elimination: 

(i) {\em spiral} hexagonal clusters, in which the outer shell is eroded sequentially in an anticlockwise manner starting from the lowest point;

(ii) {\em corner} hexagonal clusters, in which the corners of the outer shell are first removed and the erosion proceeds from all of the six corners.

(iii) {\em topdown} hexagonal clusters, in which the highest bubble in the outer shell is removed.

\noindent{\bf Hybrid cluster}: To create a cluster that is intermediate between a circular cluster and a hexagonal cluster, improving upon the method given by \citet{Coxg03}, we start from a perfect hexagonal cluster (with $N$ a hexagonal number) and remove the bubbles farthest from the centre of the cluster. This process stops when we reach another hexagonal number. (A related procedure, which makes a dodecagonal cluster by removing bubbles farthest from the centre of the cluster parallel to a line joining it to each of the six apices of the hexagonal cluster, gave similar results but with slightly greater perimeters than this one.)

\begin{figure}
\centerline{
\includegraphics[width=0.7\textwidth]{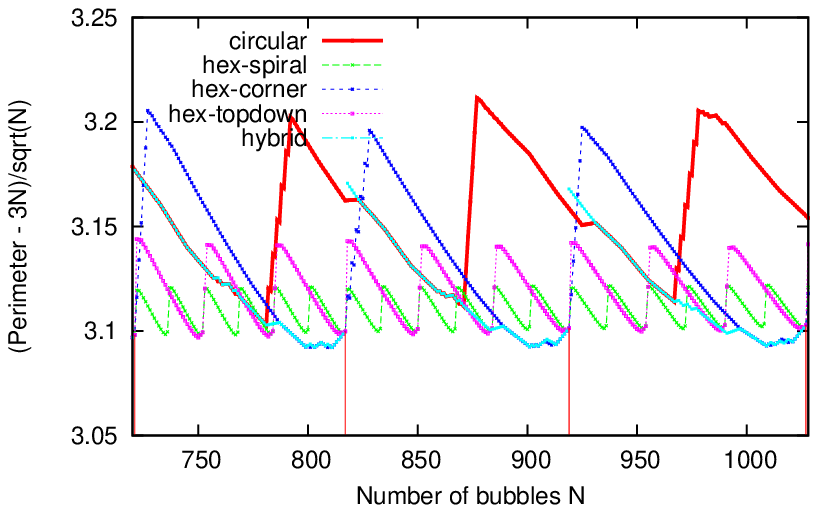}
}
\caption{Reduced perimeters for $721 \le N \le 1027$. The hexagonal numbers are marked with vertical lines.}
\label{fig:energy}
\end{figure}

\begin{figure}
\centerline{
(a)
\includegraphics[width=0.22\textwidth]{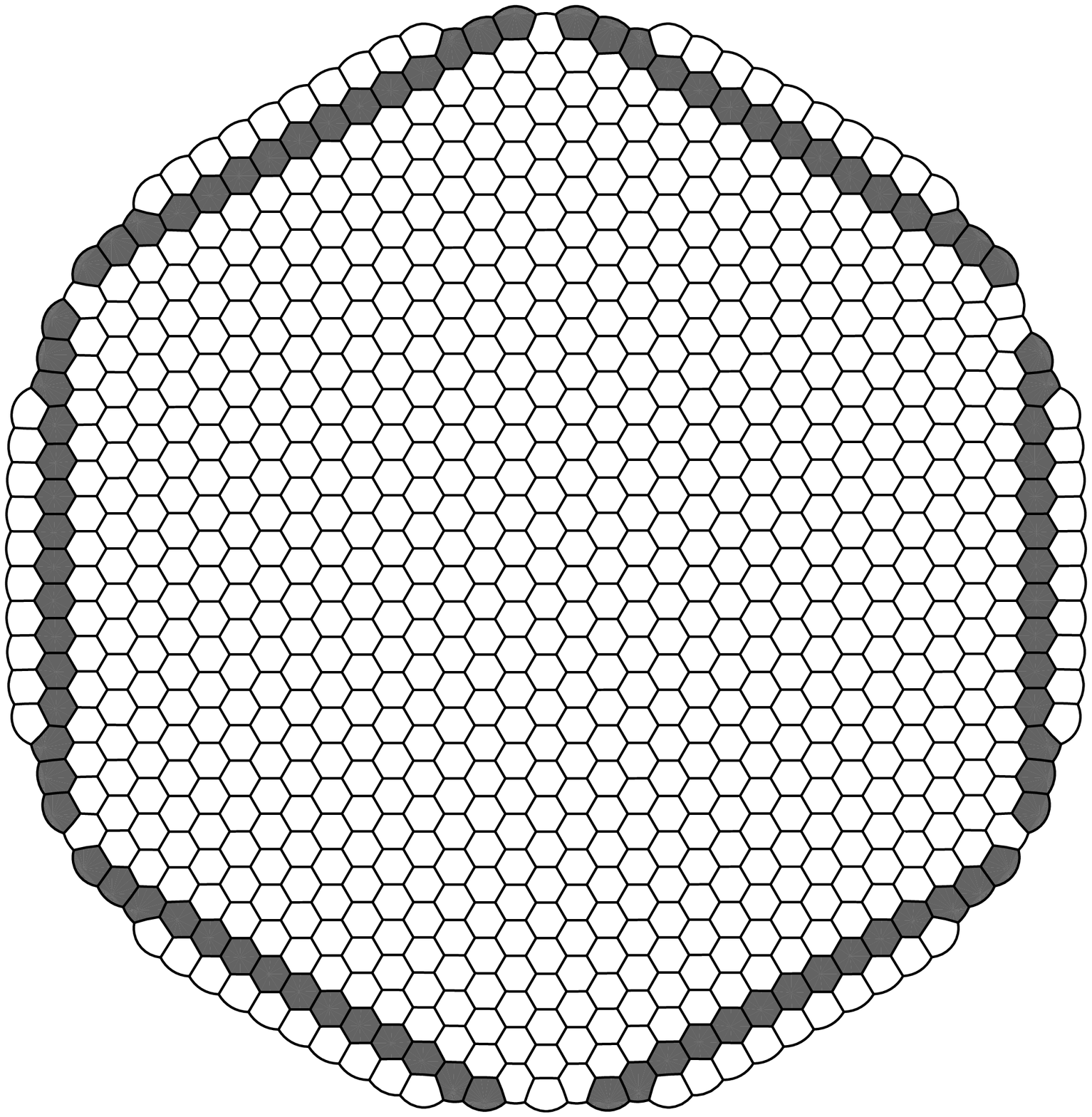}
(b)
\includegraphics[width=0.22\textwidth]{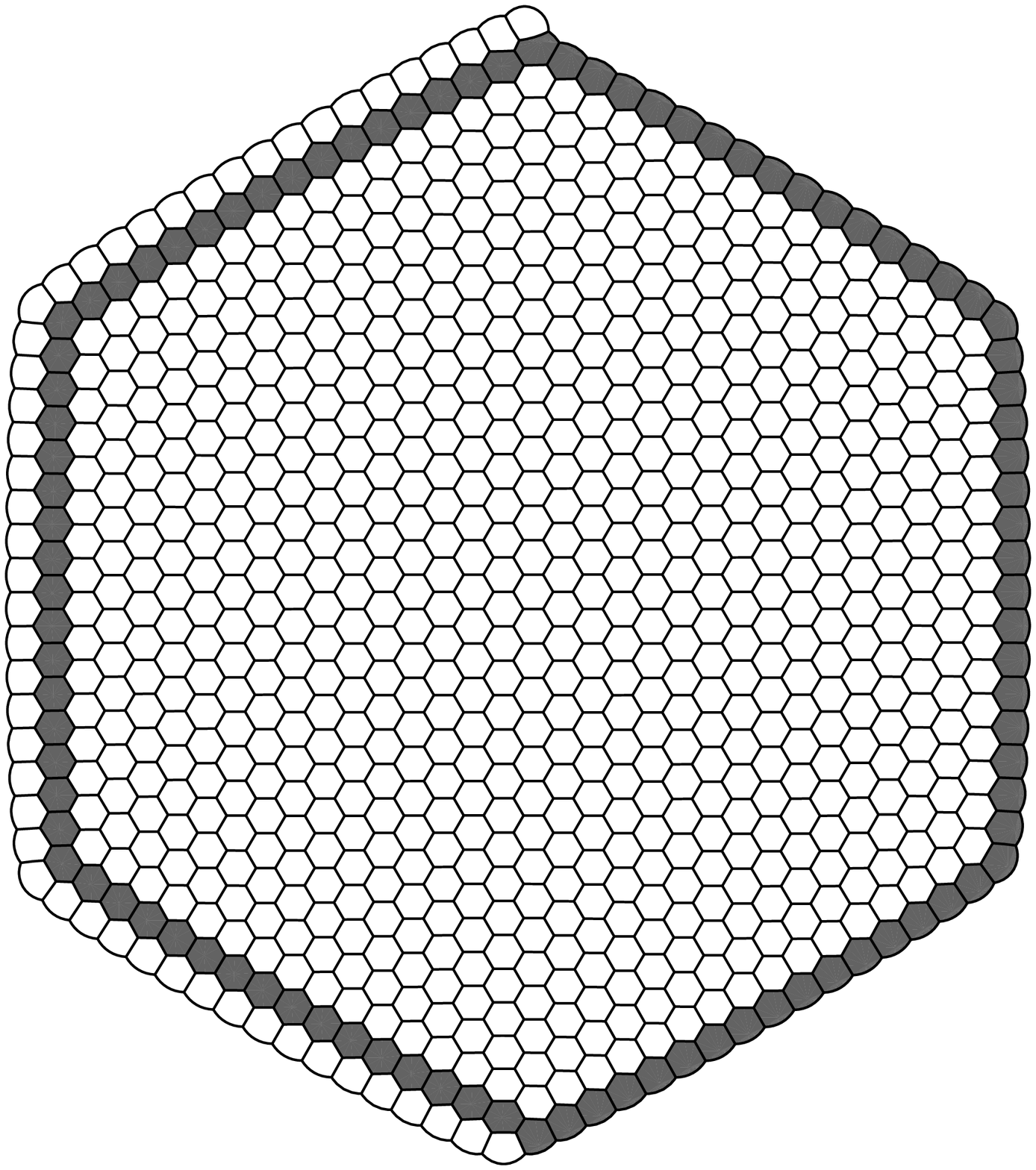}
(c)
\includegraphics[width=0.22\textwidth]{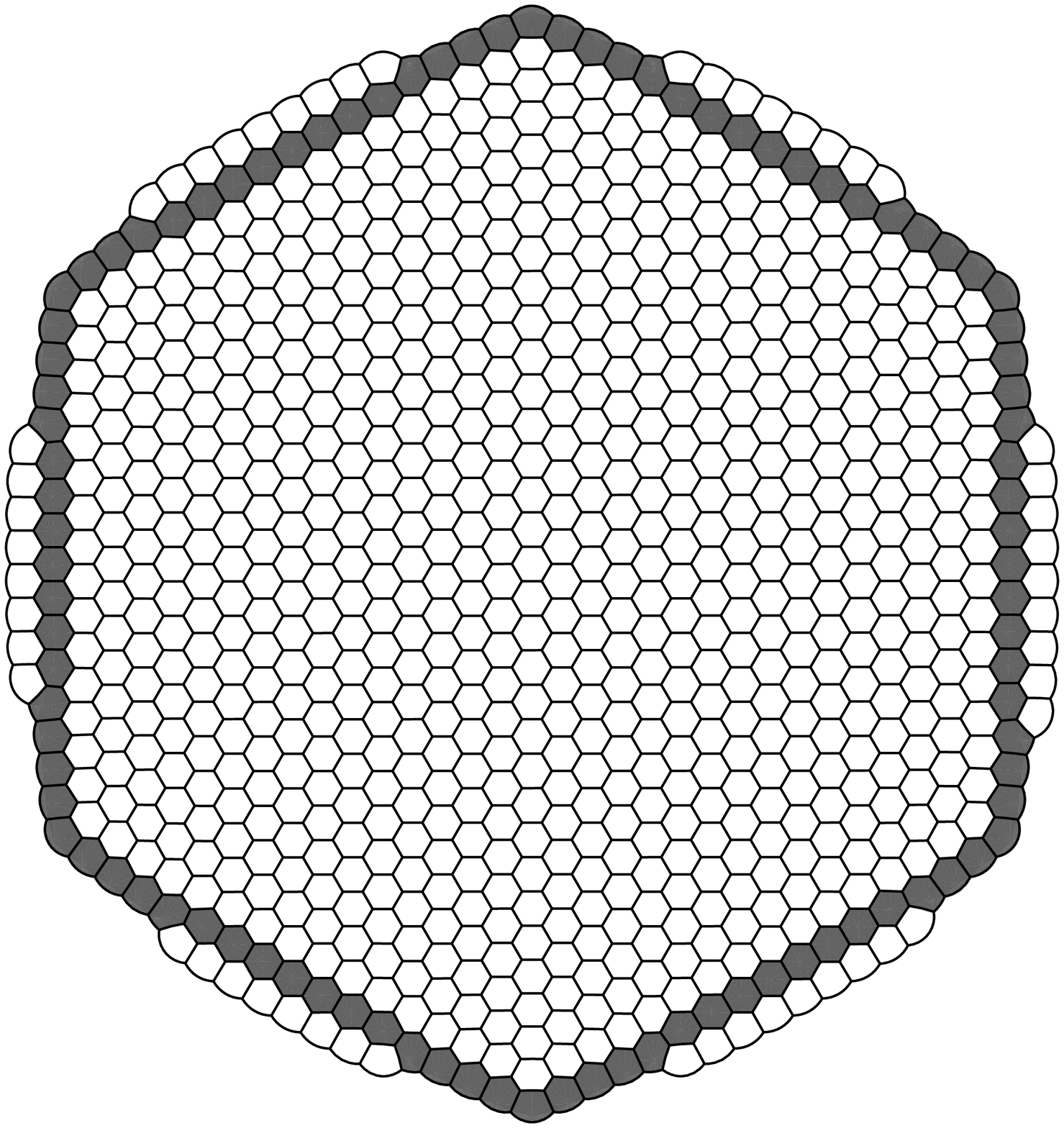}
(d)
\includegraphics[width=0.22\textwidth]{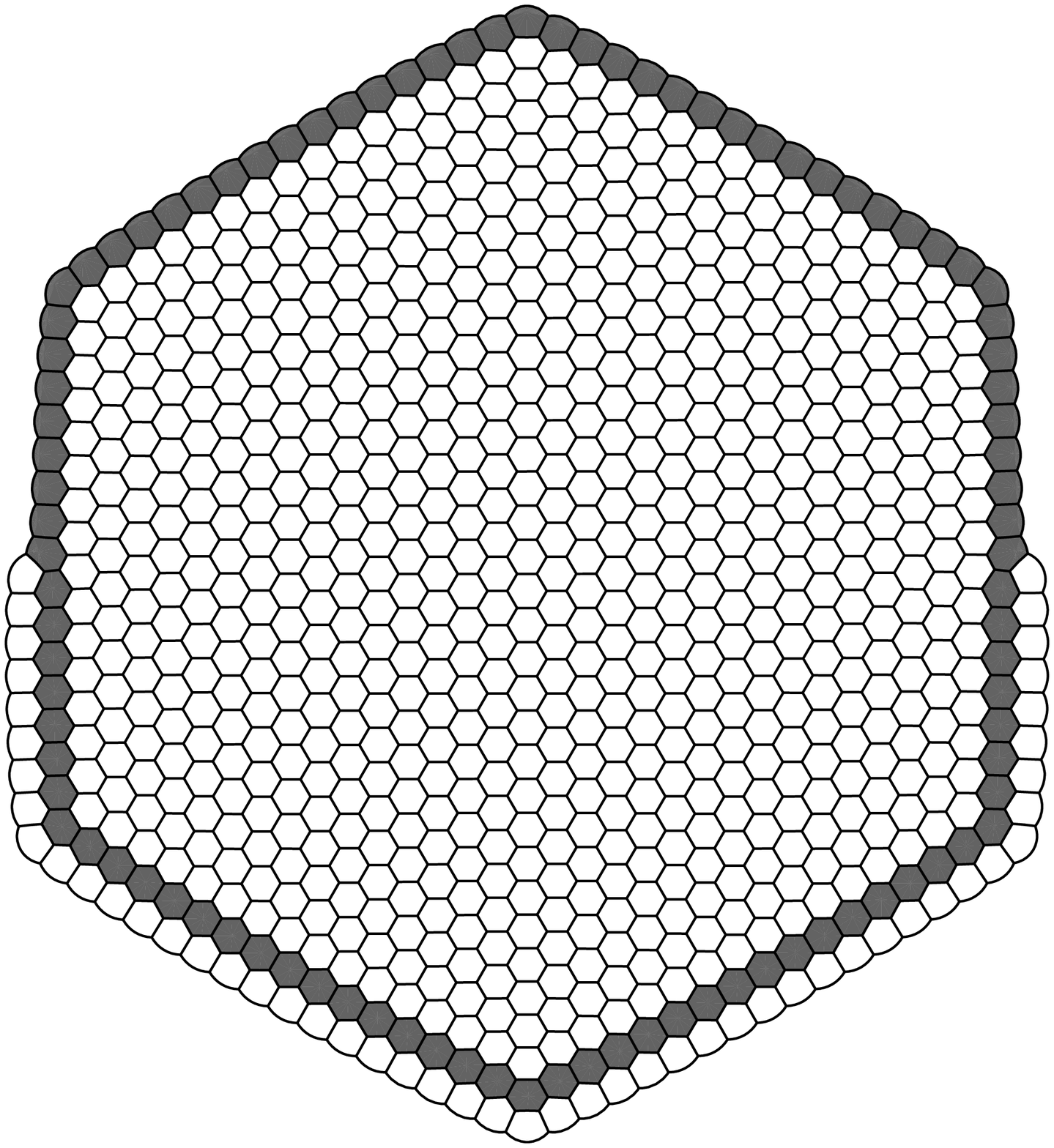}
}
\caption{Different clusters of $N=868$ bubbles, in the same order as in figure \ref{fig:hexpics}, with the penultimate shell of bubbles shaded. Circular and hybrid clusters are the same. For this $N$, the circular (or hybrid) cluster is best.}
\label{fig:868pics}
\end{figure}

\section{Results}

\subsection{Comparison of methods for $N < 1000$}

The perimeters increase approximately as $P\sim 3N+k \sqrt{N}$, with $k \approx 3.1$ \cite[]{granervmpc01}. Note that for each value of $N$ they are all close  (figure \ref{fig:hexpics}). So in figure \ref{fig:energy} we instead use what we call the {\it reduced} perimeter, $\hat{P} = (P-3N)/\sqrt{N}$. This quantity fluctuates in a saw-tooth fashion as $N$ varies, but within rather narrow limits.

The best proven general bounds on the reduced perimeter \cite[]{heppesm05} are
\begin{equation}
\sqrt{\pi A_0} - 1.5 < \hat{P} <  \pi + \frac{3}{\sqrt{N}},
\end{equation}
where the first expression is approximately 1.36. Different 
asymptotic estimates of about 3 are given by \cite{granervmpc01} and \cite{heppesm05}.

The reduced perimeter shows the greatest fluctuation as $N$ increases, with sharp upward jumps that occur roughly midway between hexagonal numbers and then a slower decay. So we should only expect that circular clusters might have the lowest perimeter far from hexagonal numbers, e.g. for $N=868$ which is midway between the hexagonal numbers $817$ and $919$ (figure \ref{fig:868pics}). 

The spiral hexagonal clusters shows six cycles in  $\hat{P}$ between hexagonal numbers, making this the cluster shape that is most likely to be best, since it shows the smallest deviations from a line joining the perimeters of the perfect hexagonal clusters. In fact, it appears to be best about a third of the time.

The topdown hexagonal clusters show three cycles between hexagonal numbers, and turn out to be better than a spiral hexagonal cluster for half the time. This shape becomes expensive when there is a half-row of hexagons along one side of the cluster, an observation that also applies to the corner hexagonal clusters. In this case, removing a small number of bubbles from all six corners of the outer shell of a hexagonal cluster is good for $N$ slightly below a hexagonal number, but this method becomes more expensive as the number removed increases because of the number of partial lines of hexagons in the outer shell. The reduced perimeter is similar to the circular case, in that it shows just one cycle between hexagonal numbers, but here the upward jump occurs for $N$ just above a hexagonal number.

\begin{figure}
\centerline{
(a)
\includegraphics[width=0.25\textwidth]{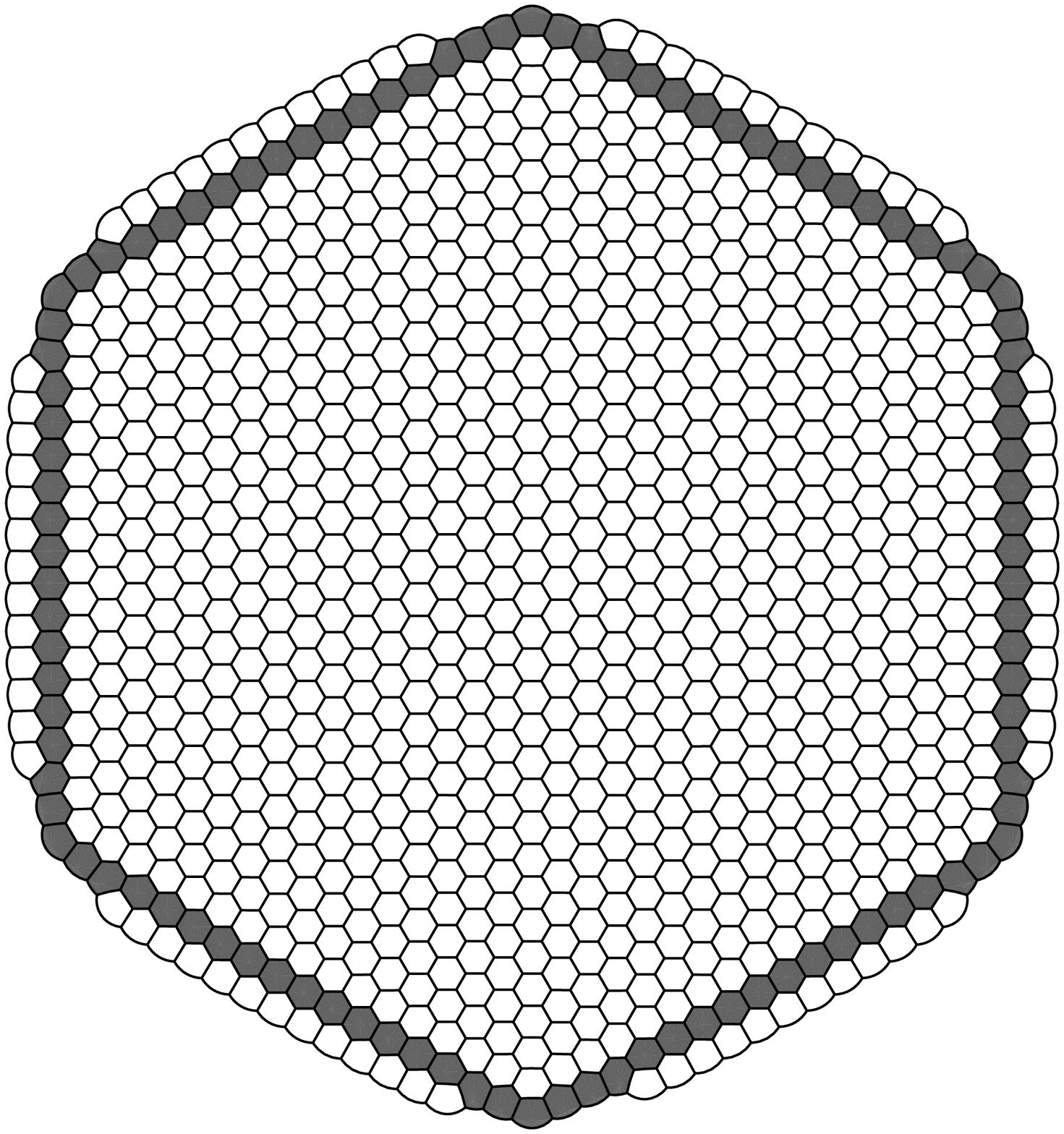}
(b)
\includegraphics[width=0.25\textwidth]{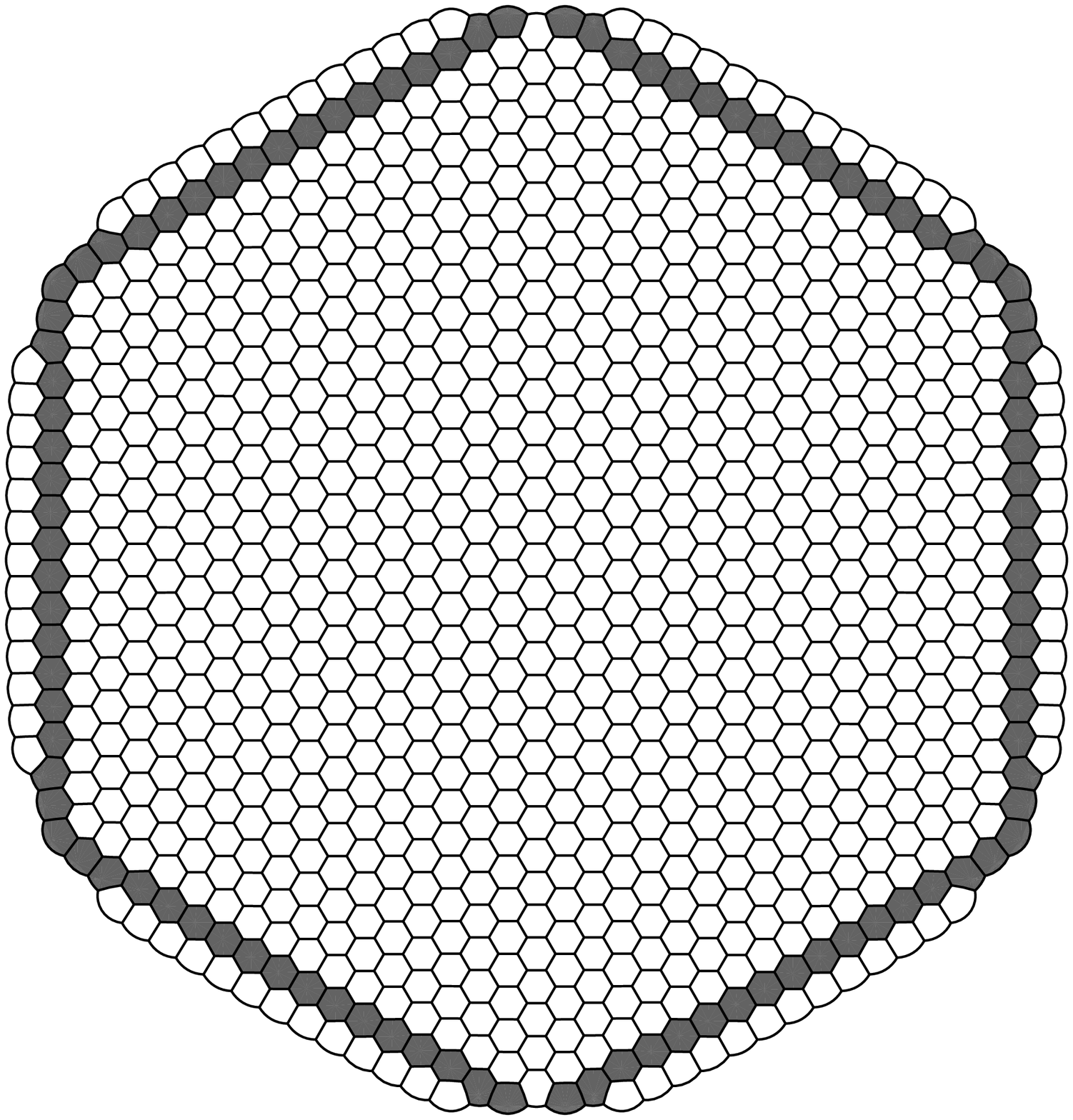}
}
\caption{Different clusters of $N=995$ bubbles, showing that for $N$ not far below a hexagonal number, the corner hexagonal and hybrid clusters are very similar, and that the hybrid cluster sometimes has slightly lower perimeter.  (a) $P_{hexc}=3082.891$. (b) $P_{hyb}=3082.799$.}
\label{fig:1000pics}
\end{figure}

A hybrid cluster is very similar to a circular cluster for $N$ less than about $200$, and to a corner hexagonal cluster (figure \ref{fig:1000pics}) for $N$ just below a hexagonal number. The difference is that after removing a few bubbles from each apex of the hexagonal cluster, the hybrid procedure allows us to remove a bubble from the next shell in. For $N$ far from a hexagonal number this method is heavily penalised: it is close to a circular cluster, and the optimal cluster is (topdown) hexagonal. As $N$ increases towards a hexagonal number, there is a short interval in which a hybrid cluster can become marginally  {\it better} than a hexagonal cluster, before the perimeter is again equal to the value in the corner hexagonal case.

\begin{figure}
\centerline{
\includegraphics[width=0.7\textwidth]{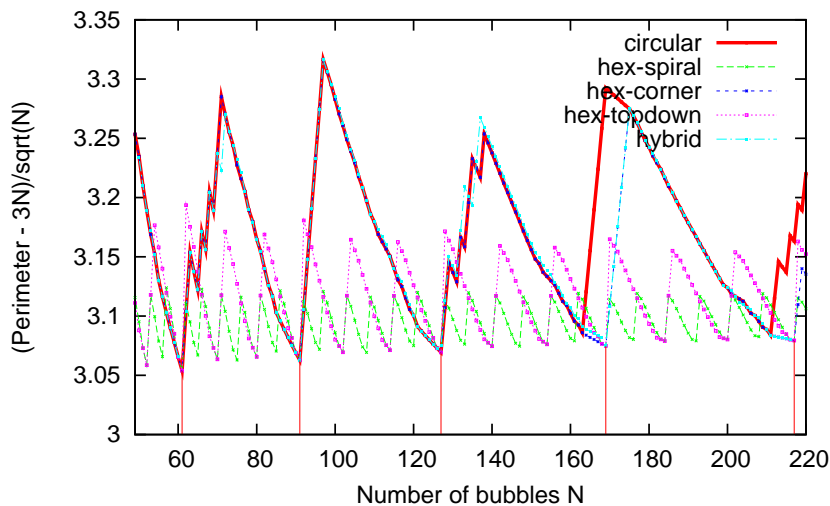}
}
\caption{Reduced perimeters for $50 \le N \le 217$. The hexagonal numbers are marked with vertical lines.}
\label{fig:energy2}
\end{figure}

This pattern is more difficult to see at low $N$, as shown in figure \ref{fig:energy2}, since circular, corner hexagonal, and hybrid clusters are often identical. This data agrees with the candidate structures for $N=50$ (topdown hexagonal) and $N=200$ (topdown or spiral hexagonal, which are equivalent here) given by \citet{Coxg03}. For $N=100$ it suggests a new candidate, adding a further defect to the periphery of the structure conjectured previously but reducing the perimeter from $330.880$ to $330.801$ (also topdown hexagonal); the result is shown in figure \ref{fig:pic100}(a).

\subsection{Influence of asymmetry}

\begin{figure}
\centerline{
(a)
\includegraphics[width=0.25\textwidth]{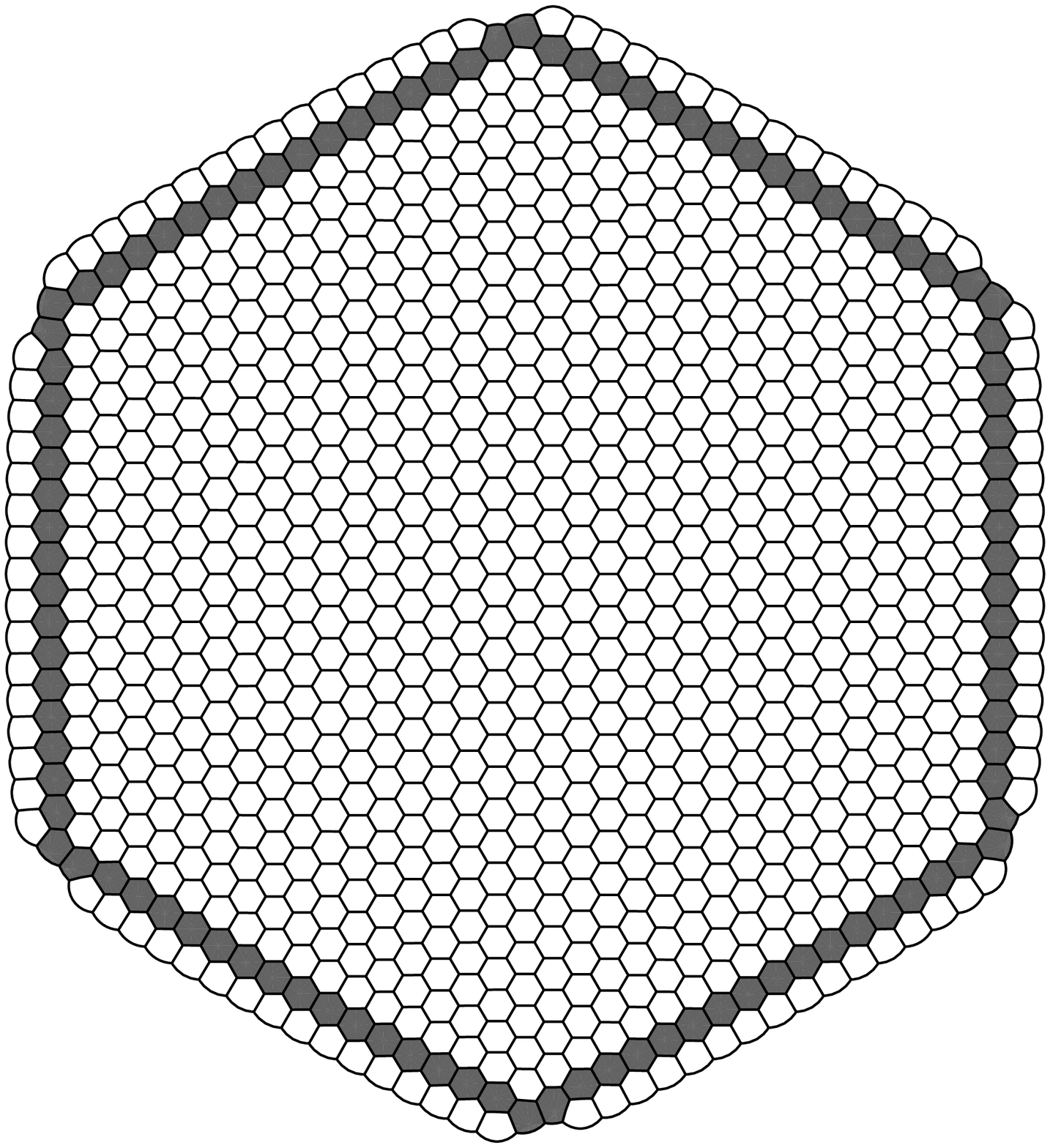}      
(b)
\includegraphics[width=0.25\textwidth]{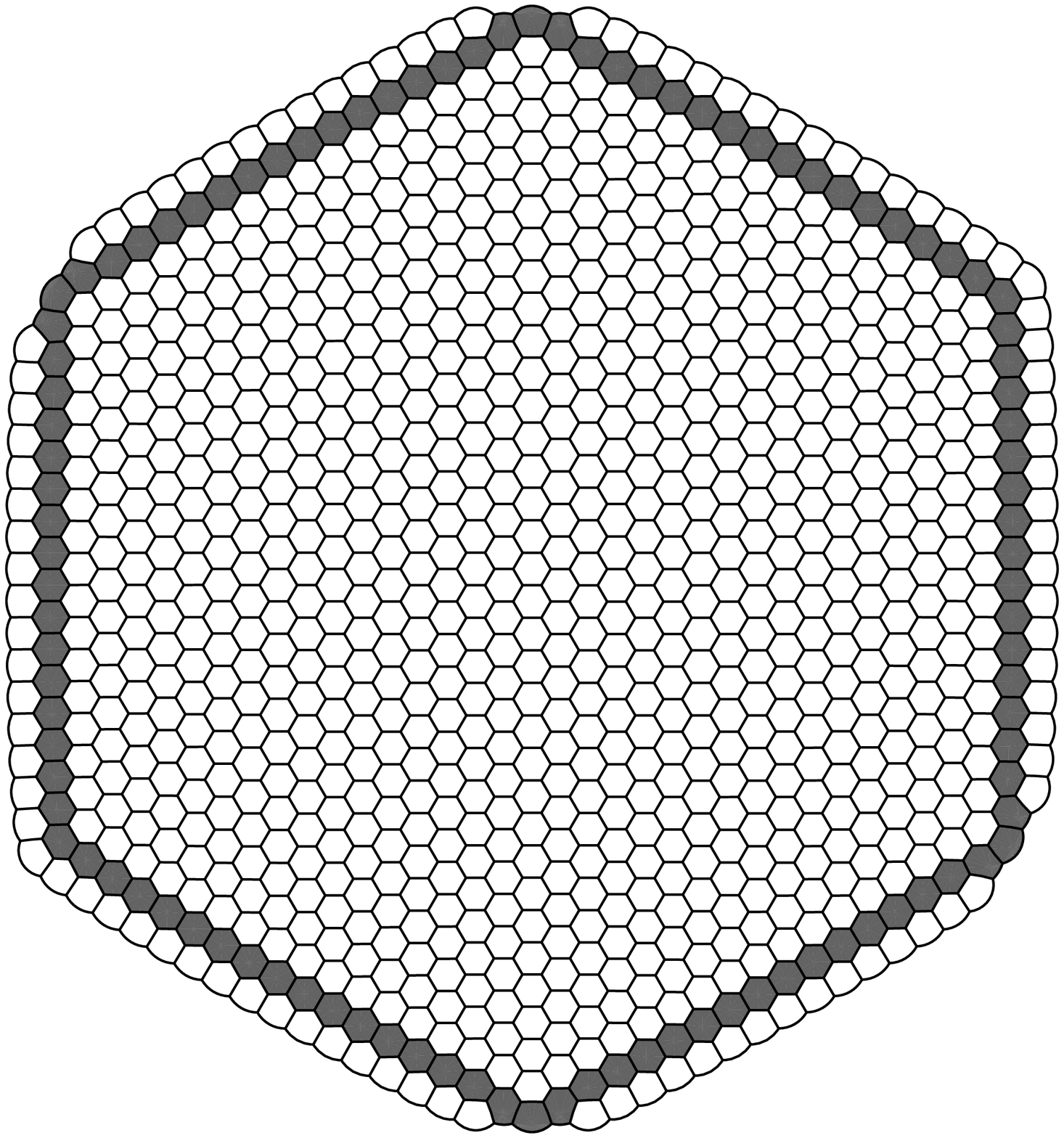}
(c)
\includegraphics[width=0.25\textwidth]{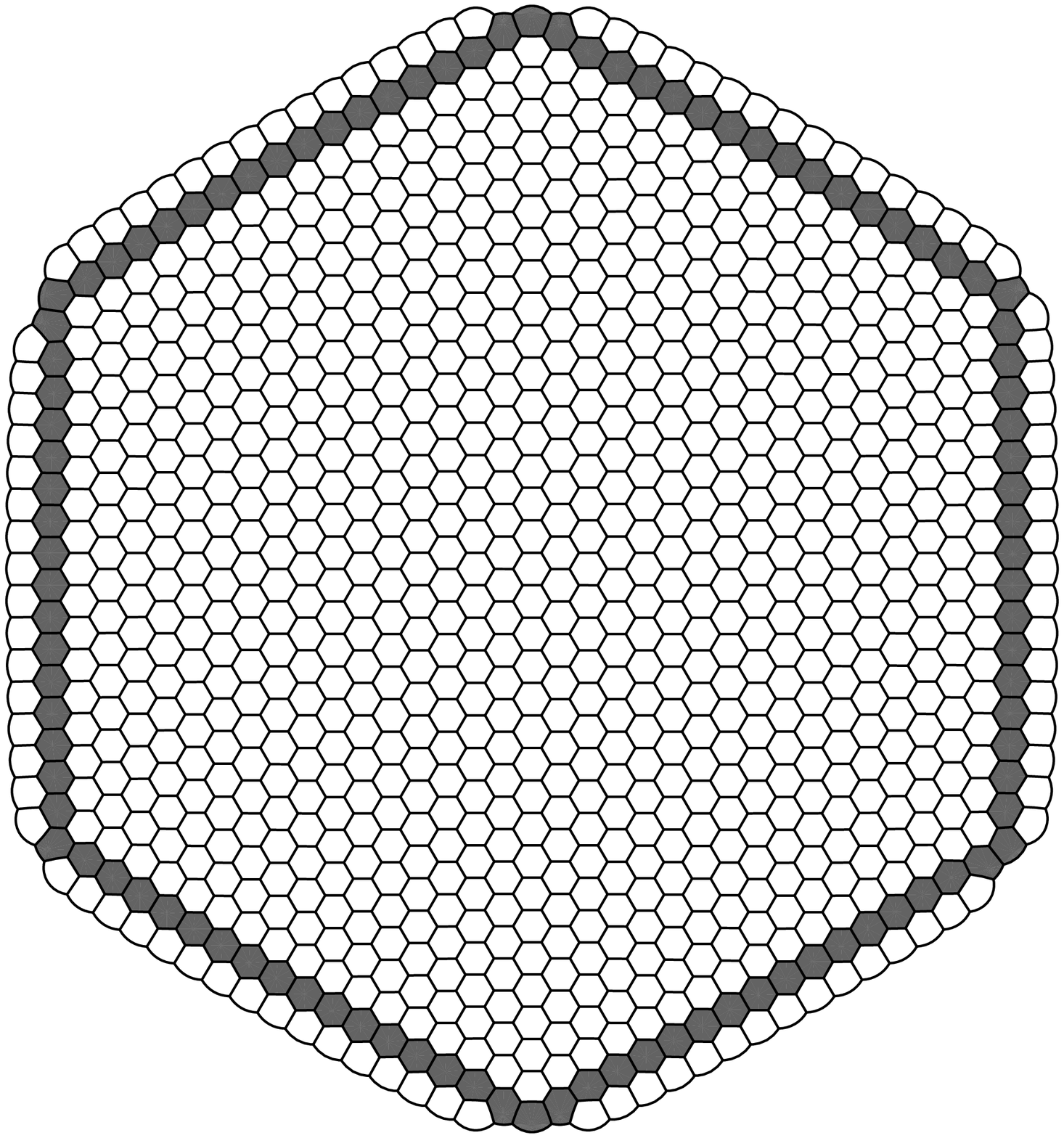}
}
\caption{Three clusters of $N=1015$ bubbles, showing that removing 12 bubbles to make a corner hexagonal cluster leads to a lower perimeter if done asymmetrically. (a) Removing two bubbles from each vertex yields perimeter $P=3143.700$. (b) Removing three bubbles from each of four vertices yields perimeter $P=3143.643$. (c) Removing three bubbles from a pair of vertices, two from another pair, and one from the third pair yields the lowest perimeter $P=3143.613$. }
\label{fig:1015}
\end{figure}

There is also a small discrepancy in the data, visible in figure \ref{fig:energy}, that turns out to be significant. For $N$ just below a hexagonal number, the corner hexagonal clusters and the hybrid cluster are slightly different, although the methods described above should give exactly the same answer. The discrepancy is due to the way in which bubbles are removed in a corner hexagonal cluster: perhaps because of the definition of cluster ``centre'', not all apices are treated equally. Figure \ref{fig:1015} shows three different clusters of $N=1015$ bubbles, which is twelve less than the hexagonal number 1027. Instead of removing two bubbles from each corner, we instead remove 3 from two corners, and 1 from two corners. This asymmetric cluster turns out to have lower perimeter!

\begin{figure}
\centerline{
(a)
\includegraphics[width=0.25\textwidth]{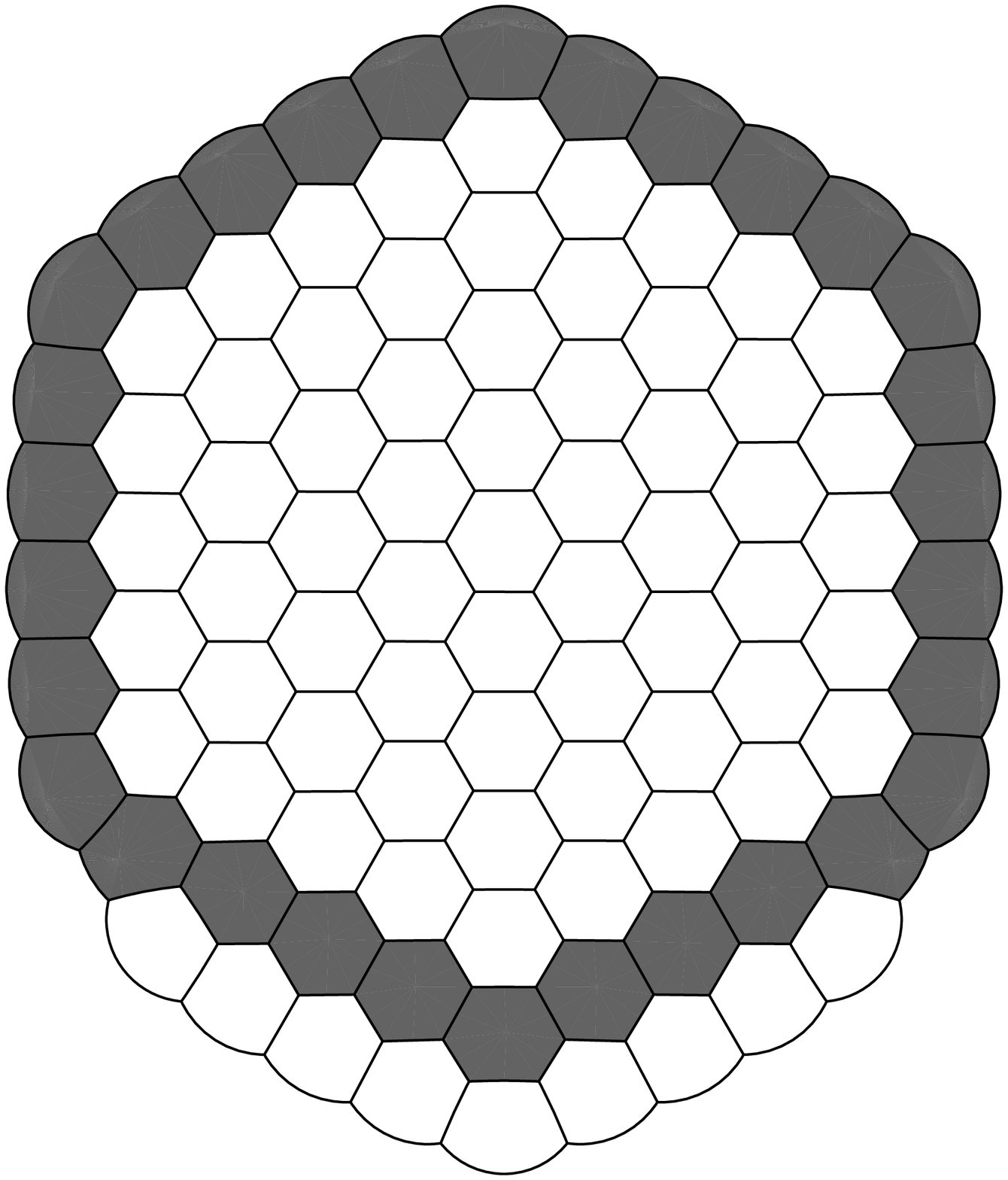}      
(b)
\includegraphics[width=0.25\textwidth]{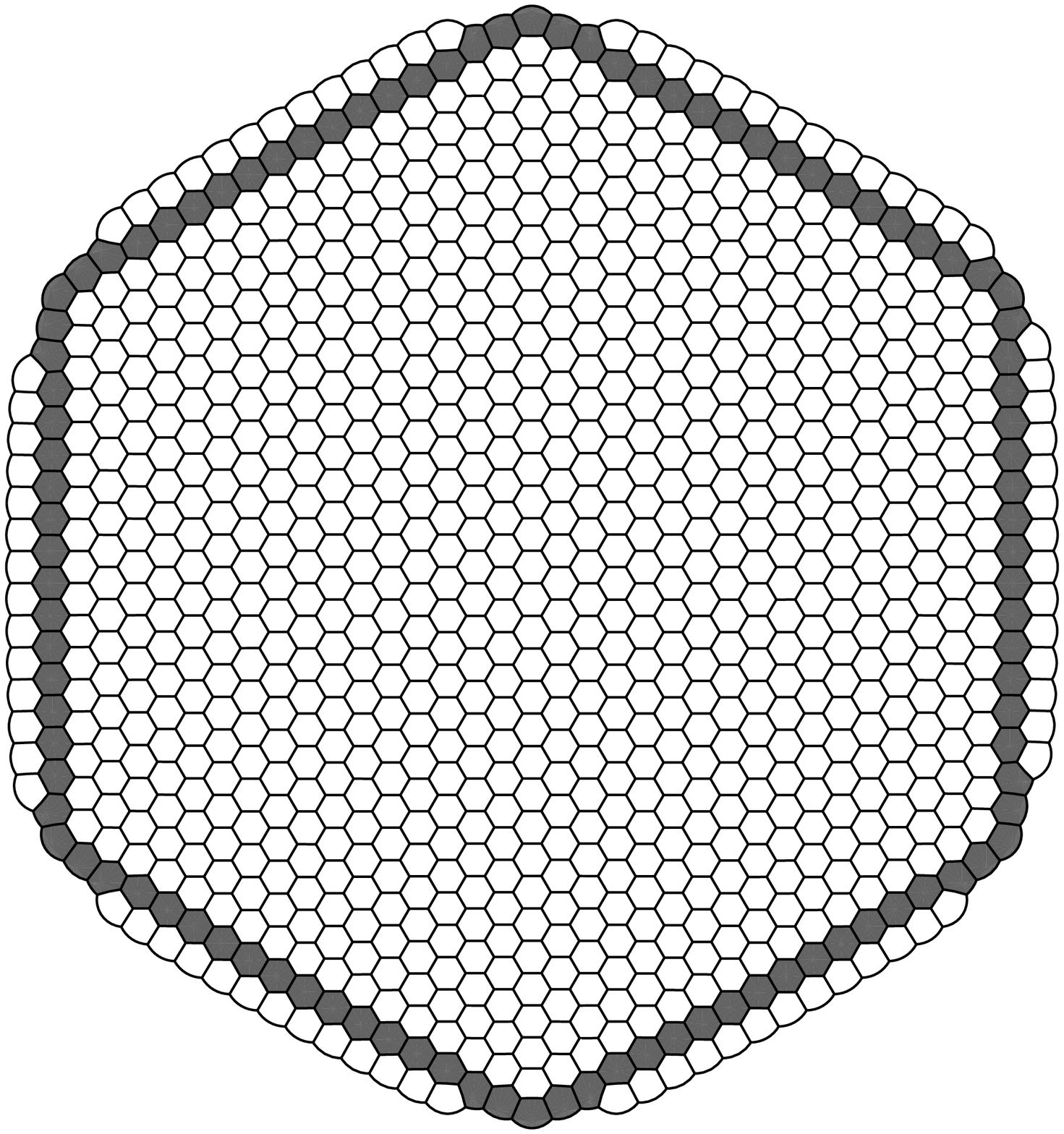}
(c)
\includegraphics[width=0.25\textwidth]{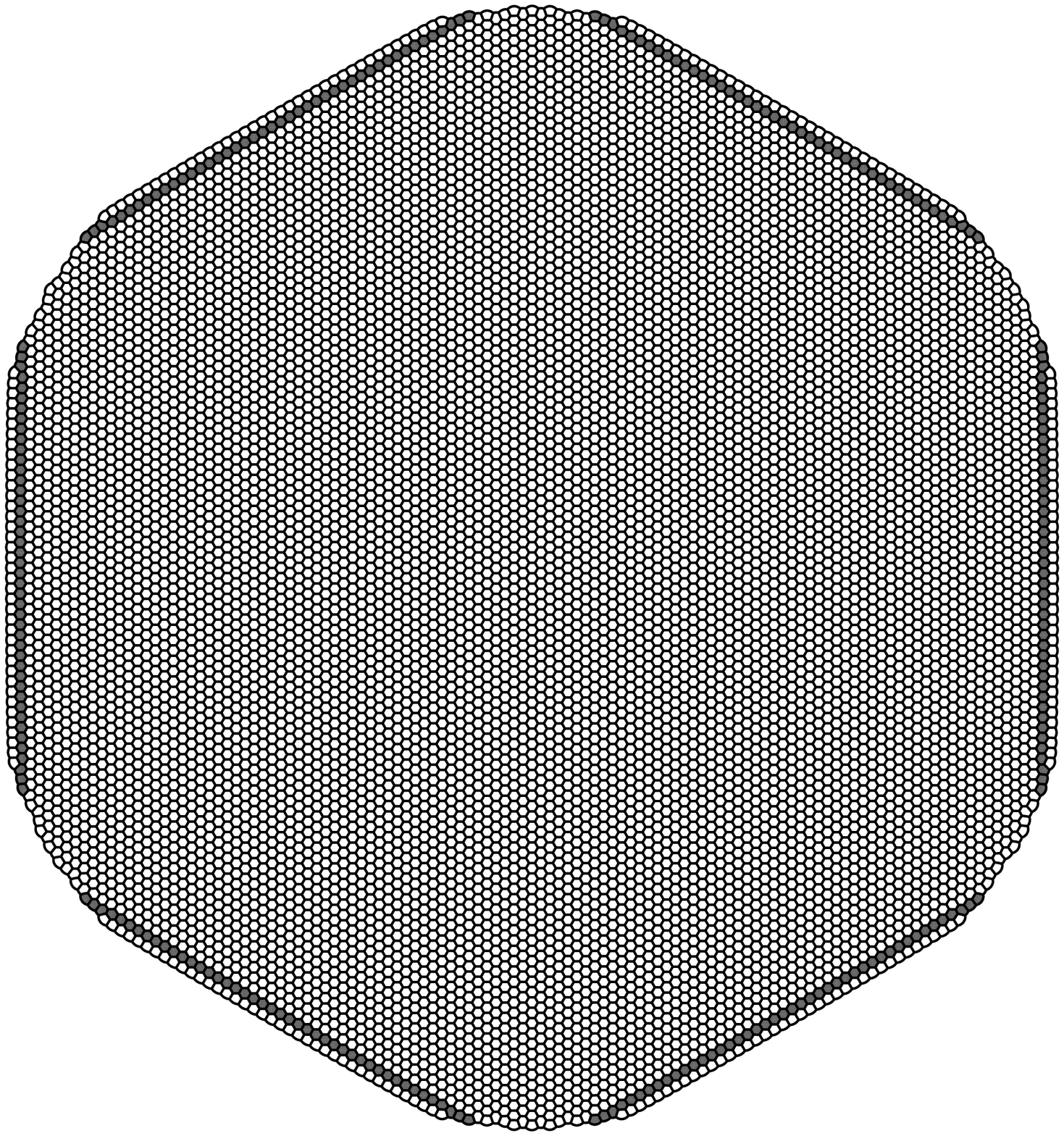}
}
\caption{New candidate minimal clusters (a) $N=100$ bubbles with perimeter $P_{hext}=330.801$. 
(b) $N=1000$ bubbles with perimeter $P_{hexc}=3098.003$.
(c) $N=10,000$ bubbles with perimeter $P_{hyb}=30310.532$.
}
\label{fig:pic100}
\end{figure}

In particular, this allows us to suggest a new candidate configuration for the optimal cluster of $N=1000$ bubbles, shown in figure \ref{fig:pic100}(b), which is an asymmetric corner hexagonal cluster.

\subsection{When $N$ is a large hexagonal number}

$N=1,000$ is too small for any rounding of the corners of a hexagonal cluster to reduce its total perimeter. We consider $N=10,000$, and find that a hybrid cluster constructed by removing bubbles from the hexagonal cluster of $N=10,267$ {\em does} beat all possible hexagonal candidates described here: this candidate for $N=10,000$ has $P_{hyb} = 30310.532$ compared to the best hexagonal case (topdown hexagonal) with $P_{hext} =30312.589$. 
 
\begin{figure}
\psfrag{Perimeter - P_{hex}}{\small Perimeter - $P_{hex}$}
\centerline{\includegraphics[width=0.7\textwidth]{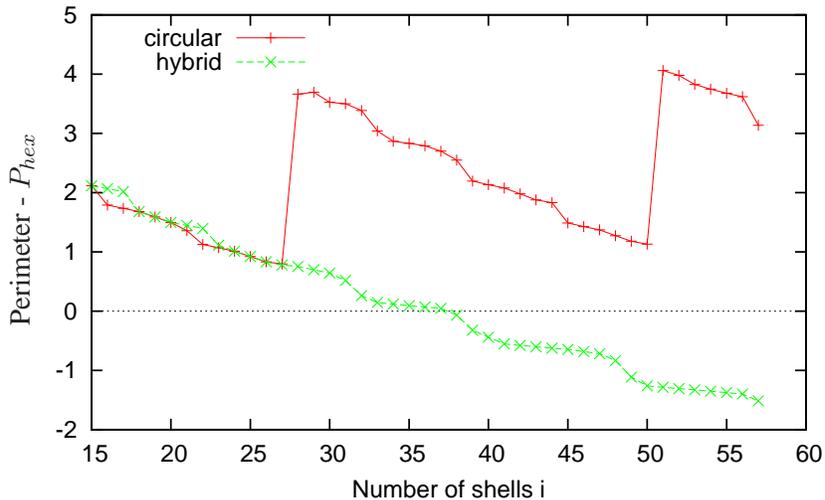}}
\caption{Difference between the perimeter of a circular cluster or a part-circular cluster and a perfect hexagonal cluster for $N$ of the form $3 i^2 + 3i +1$, where $i$ is the number of shells of hexagons in the perfect cluster. For $i \ge 38$, corresponding to $N \ge 4447$ bubbles, the part-circular cluster has lower perimeter, while a circular cluster never beats a hexagonal cluster.}
\label{fig:energy3}
\end{figure}

This suggests that, for sufficiently large $N$, even for a hexagonal number the pure hexagonal cluster may not be best, as Morgan conjectures. We therefore use the hybrid method to reduce each hexagonal cluster until $N$ reaches the next hexagonal number, and compare the perimeter  with the perfect hexagonal one. Figure \ref{fig:energy3} shows that for $N\ge 4447$ the hybrid cluster becomes better than the hexagonal cluster for a hexagonal number $N$, and the resulting best perimeters are recorded in Table \ref{tab:hexN}. 

Why does the crossover occur at $N=4447$? As $N$ increases, the hybrid method leaves a higher proportion of peripheral bubbles from the hexagonal cluster from which it was constructed (and therefore more bubbles are removed from inner shells). That is, for $N$ greater than about 3500 the proportion of bubbles in the outer layer of the hybrid cluster that were in the outer shell of the hexagonal cluster exceeds 0.5. For  larger $N$, the cluster becomes more ``dodecahedral'' than circular, and this appears to reduce the perimeter.

\begin{table}
\begin{center}
\begin{tabular}{c|c||c|c||c|c}
$N$ & $P$ & $N$ & $P$& $N$ & $P$ \\ \hline
721	&	2246.190&	2791	&	8537.503&	6211	&	18878.290		\\
817	&	2539.604&	2977	&	9100.929&	6487	&	19711.685		\\
919	&	2851.019&	3169	&	9682.356&	6769	&	20563.080		\\
1027	&	3180.435&	3367	&	10281.783&	7057	&	21432.396		\\
1141	&	3527.852&	3571	&	10899.211&	7351	&	22319.550		\\
1261	&	3893.271&	3781	&	11534.639&	7651	&	23224.833		\\
1387	&	4276.691&	3997	&	12188.067&	7957	&	24148.243		\\
1519	&	4678.112&	4219	&	12859.496&	8269	&	25089.652		\\
1657	&	5097.533&	4447	&	13548.857&	8587	&	26049.062		\\
1801	&	5534.955&	4681	&	14256.032&	8911	&	27026.473		\\
1951	&	5990.378&	4921	&	14981.347&	9241	&	28021.883		\\
2107	&	6463.802&	5167	&	15724.660&	9577	&	29035.292		\\
2269	&	6955.226&	5419	&	16486.067&	9919	&	30066.610		\\
2437	&	7464.651&	5677	&	17265.473	&&				\\
2611	&	7992.077&	5941	&	18062.882	&&				\\
\end{tabular}
\end{center}
\caption{Perimeter of candidate clusters to the least perimeter arrangement of $N$ bubbles of area $3\sqrt{3}/2$ for $N$ a hexagonal number between 721 and 9919.}
\label{tab:hexN}
\end{table}

\subsubsection{Extending the hybrid method}

Recall that we can use the hybrid method described in \S \ref{sec:methods} to eliminate bubbles from a hexagonal cluster to arrive at a slightly rounded cluster with a  number of bubbles that is the next lowest hexagonal number of the form $3i^2 + 3i+1$. For sufficiently large $N$ this procedure may be repeated, to arrive at a more rounded cluster for the next lowest hexagonal number. In the limit, we reach the circular case.

\begin{figure}
\psfrag{P - P_{hex}}{\small Perimeter - $P_{hex}$}
\psfrag{Initial number of additional shells}{\small Number of extra initial shells}
\centerline{
\includegraphics[width=0.7\textwidth]{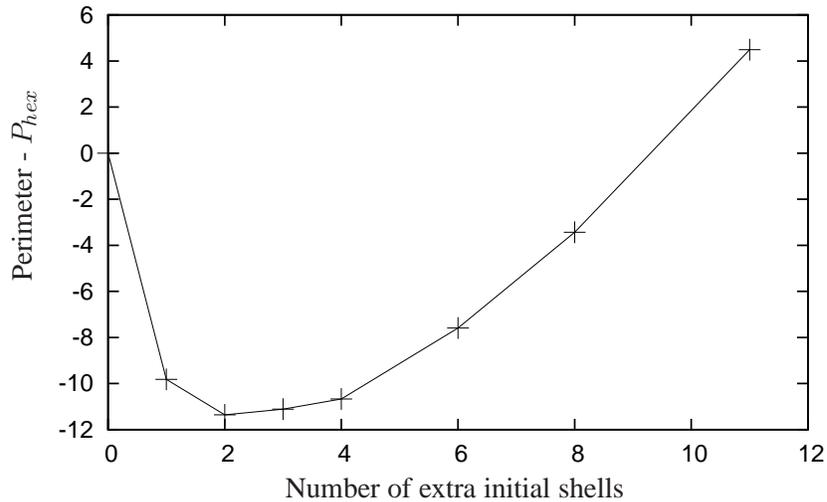}
}
\caption{Perimeter, expressed as the difference from the hexagonal cluster, of different clusters of $N=170,647$ bubbles created with a generalized hybrid procedure. The curve saturates to the right, since the circular limit is reached here.
}
\label{fig:verylarge}
\end{figure}

To illustrate this, we choose the value $N=170,647$ ($i=238$) to compare the effect of starting the hybrid procedure from different hexagonal clusters. For this $N$, the hexagonal cluster has $P_{hex} = 513,236.338$ and a circular cluster has greater perimeter, $P_{circ}= 513,240.830$. A hybrid cluster created from $N=172,081$ in the usual way has even lower perimeter, $P_{hyb} = 513,226.522$, but starting from $N=176,419$ and removing the furthest 5772 bubbles from the centre gives a cluster with an even lower perimeter, $P_{hyb2} = 513,224.982$. This result is shown in figure \ref{fig:verylarge}, suggesting that the global minimum is found when the procedure starts from a hexagonal cluster that is two shells larger than required (so the minimum in the number of layers removed presumably increases very slowly with $N$). Note that the difference in perimeter is a small fraction of the total. Note also that for such large clusters, the energy minimisation (gradient descent) in Surface Evolver takes around 3 days for each cluster.

\section{Conclusion}

For $N$ less than about one thousand, the sequence of optimal observed clusters, increasing from one hexagonal number to the next, is corner hexagonal, spiral hexagonal, corner hexagonal, spiral hexagonal, possibly hybrid or circular, then topdown hexagonal, hybrid and back to corner hexagonal. 
For larger $N$, the optimal cluster is less likely to be hexagonal in shape, even for $N$ a hexagonal number, and we find that for $N \ge 4447$ the perfect hexagonal cluster is no longer best even for a hexagonal number of bubbles.

It is clear that for each $N$, there are still many possible small changes to each sort of cluster that could be tried in seeking a better minimum. One possibility is to extend our definition of hexagonal to allow more than one layer of bubbles to be shaved off any one of the six sides of the cluster. Another is to exploit the apparent improvement with the introduction of a little asymmetry, as illustrated in figure \ref{fig:1015}. 

It also remains to determine if the limiting behavior of a perimeter-minimizing 
cluster of $N$ equal-area bubbles as $N$ approaches infinity is circular.

\subsection*{Acknowledgements}

SJC and FM acknowledge the support of the ICMS during the workshop ``Isoperimetric problems, space-filling, and soap bubble geometry''. We thank K. Brakke for developing and distributing the Surface Evolver, and SJC thanks EPSRC (EP/D071127/1) for funding.


\end{document}